# Multi-resolution community detection based on generalized self-loop rescaling strategy

Ju Xiang[1], Yan-Ni Tang[1], Yuan-Yuan Gao[1], Yan Zhang[2,*], Ke Deng[3], Xiao-Ke Xu[4] and Ke Hu[5,*]

[1]*Department of Basic Medical Sciences, Changsha Medical University, Changsha 410219, Hunan, The People's Republic of China*
[2]*Department of Computer Science, Changsha Medical University, Changsha 410219, Hunan, The People's Republic of China*
[3]*Department of Physics, Jishou University, Jishou 416000, Hunan, The People's Republic of China*
[4]*School of Communications and Electronic Engineering, Qingdao Technological University, Qingdao 266520, Shandong, The People's Republic of China*
[5]*Department of Physics, Xiangtan University, Xiangtan 411105, Hunan, The People's Republic of China*

**Abstract:**
Community detection is of considerable importance for analyzing the structure and function of complex networks. Many real-world networks may possess community structures at multiple scales, and recently, various multi-resolution methods were proposed to identify the community structures at different scales. In this paper, we present a type of multi-resolution methods by using the generalized self-loop rescaling strategy. The self-loop rescaling strategy provides one uniform ansatz for the design of multi-resolution community detection methods. Many quality functions for community detection can be unified in the framework of the self-loop rescaling. The resulting multi-resolution quality functions can be optimized directly using the existing modularity-optimization algorithms. Several derived multi-resolution methods are applied to the analysis of community structures in several synthetic and real-world networks. The results show that these methods can find the pre-defined substructures in synthetic networks and real splits observed in real-world networks. Finally, we give a discussion on the methods themselves and their relationship. We hope that the study in the paper can be helpful for the understanding of the multi-resolution methods and provide useful insight into designing new community detection methods.

## 1. Introduction

Community detection in complex networks attracted much attention from various fields. Many complex networks, including social, biological and technological networks, consist of communities—groups of vertices with dense internal connections and sparser external connections [1]. Detecting and analyzing such communities in the networks is of considerable importance for understanding the structure and function of the networks as well as dynamics taking place on the networks [2-6]. In recent years, many community-detection algorithms have been proposed based on various approaches [7-12] (see Refs [1, 13, 14] for reviews). Particularly, the optimization of modularity—a quality function for community partition of network proposed by Newman and Girvan—becomes one of the most popular strategies for community detection [15-21]. Nevertheless, the modularity optimization also encounters some difficulties [22, 23]. Especially, it can only produce the community structure at a certain characteristic scale, known as the resolution limit of modularity [24]. However many real-world networks may possess community structures at multiple scales [1].

Recently, various multi-resolution methods were proposed and applied to the analysis of complex networks, which can easily circumvent the resolution limit and help in discovering multi-scale structures in networks [25, 26]. For example, multi-resolution modularity methods with tunable parameters have been studied [25, 27, 28]. Statistical clustering methods in data mining have been applied to the analysis of multi-scale structure in networks [29, 30]. Some multi-resolution methods are based on the correlation between dynamics and multi-scale structures in networks [31-35]. Some methods make use of the local

---

optimization of fitness functions [36, 37]. Some methods are based on Potts model [38-41]. In Ref [42], the time of dynamical processes can be used as a resolution parameter.

In this paper, we propose a type of multi-resolution methods based on a kind of generalized self-loop rescaling strategy. We provide one uniform ansatz for the design of multi-resolution methods by the generalized self-loop rescaling strategy. Many quality functions for community detection can be unified in the framework of the self-loop rescaling. The resulting multi-resolution modularity by the self-loop rescaling strategy can be optimized directly using the existing modularity-optimization algorithms with minimum code development, because it is only indirectly that the modularity is modified by using this strategy. We apply the multi-resolution methods to several synthetic and real-world networks. The results show that these methods can find the predefined substructures in synthetic networks and exact splits of real networks reported in the literatures. Finally, we discuss the relationship of these multi-resolution methods with other methods and related problems.

## 2. Multi-resolution methods based on self-loop rescaling

In this section, we will describe the Newman-Girvan modularity. Then, by generalizing the self-loop rescaling strategy, we propose one general and uniform ansatz for the design of multi-resolution community detection methods, and several multi-resolution methods are derived from the modularity by the self-loop rescaling strategy.

*2.1. Newman-Girvan modularity*

In reference [43], Newman and Girvan proposed a quality function, *modularity Q*, to evaluate the quality of the found community partitions of networks, which allows the quantification of community structures in networks without *a priori* specifying the number of the communities and their sizes. The value of modularity is determined by the network topology, the community partition and the null models as comparison. Given a community partition of a network, the Newman-Girvan modularity can be written as

$$Q = \frac{1}{2M}\sum_{i,j}(A_{ij} - \frac{k_i k_j}{2M})\delta(C_i, C_j) = \sum_s \frac{k_s^{in}}{2M} - (\frac{k_s}{2M})^2, \quad (1)$$

where $A_{ij}$ is the adjacent matrix of the network ($A_{ij}=1$ if there exists an edge between vertices $i$ and $j$, and zero otherwise); $k_i = \sum_j A_{ij}$ is the degree of vertex $i$; $C_i$ is the group that vertex $i$ is assigned into; the Kronecker delta function $\delta(C_i, C_j)$ takes the value of 1 if vertices $i$ and $j$ are assigned into the same group, and 0 otherwise; $2M = \sum_{ij} A_{ij}$ is the total degree; $k_s^{in}$ is the inner degree of community $s$, $k_s$ is the total degree of community $s$.

The null model as comparison is crucial to the definition of modularity. The null model in Newman-Girvan modularity is called as configuration model (CM) and is usually preferred because it takes into account the degree heterogeneity of network under study. But other null models such as Erdös-Rényi null model (ER) recently also received extensive discussions [8, 40]. The modularity provides a way objectively to evaluate the quality of community partitions of networks, but it also suggests an alternative for community detection—modularity optimization, which is a popular strategy for community detection. However, the modularity optimization also suffers from some difficulties [22, 23], such as the resolution limit problem widely discussed [24], which can be expressed by $k_s k_t < 2M \cdot e_{st}$, where $k_s$ and $k_t$ are the total degrees of communities $s$ and $t$, $e_{st}$ is the number of edges between them, and $M$ is the total number of edges in the network. If the inequality is satisfied, the communities $s$ and $t$ will be assigned into one group by the modularity-based methods. The resolution limit means that the modularity optimization can only detect communities at a certain characteristic scale. This is not compatible to multi-scale structures in many real-world networks. Therefore, it is necessary to design multi-resolution methods for community detection.

*2.2. General description of self-loop rescaling strategy*

One of the simplest ways to the resolution limit of modularity is to add a resolution parameter into the definition of modularity, resulting the well-known multi-resolution modularity [8, 27, 28]. The multi-resolution modularity can be constructed directly by modifying the definition of modularity, but also can be derived indirectly from the Newman-Girvan modularity by self-loop rescaling strategy, i.e. by rescaling the network topology with a suitable *self-loop* assignment, because the self-loop rescaling strategy is able to indirectly modify the null model in modularity and its weight [25]. And, one of the advantages of the self-loop rescaling strategy is that the resulting multi-resolution modularity can be optimized by the existing modularity optimization algorithms such as the extremal optimization algorithm [15], the fast greedy algorithm of Newman [44] and the fast greedy algorithm of Blondel et al [16]. This clearly enriches the application of the modularity optimization algorithms in community detection.

In general, the mathematical definition of the multi-resolution modularity based on the self-loop rescaling strategy can be written as

$$Q^* = Q\left\{A_{ij} + \gamma k_i^{eff} \cdot I_{ij} - k_i \cdot I_{ij}\right\}$$
$$= \frac{\gamma - 1}{\gamma} + \frac{1}{\gamma}\sum_s \frac{k_s^{in}}{2M} - \gamma\left(\frac{k_s^{eff}}{2M}\right)^2, \quad (2)$$

where $\gamma$ is a tunable resolution parameter; $I_{ij}$ is the identity matrix; $k_i^{eff}$ as a function of the adjacent matrix, is the *effective* degree of node $i$, and $\sum_i k_i^{eff} = 2M$ ; $k_s^{in}$ is the inner degree of group $s$; $k_s^{eff}$ is the *effective* total degree of group $s$, which is the sum of the *effective* degree of nodes in group $s$; other notations are the same as in equation (1). The representation above means to provide each node a self-loop of strength $\gamma k_i^{eff} - k_i$ so as to rescale the network topology. The constant term before the summation has no effect on the $Q^*$ optimization at a given value of $\gamma$, while it is the necessary result of the self-loop strategy. As we see, by the self-loop rescaling strategy, the inequality of resolution becomes $\gamma k_s^{eff} k_t^{eff} < 2M \cdot e_{st}$. Clearly, the resolution of the modularity with self-loop rescaling can be affected by the resolution parameter $\gamma$ and the form of the *effective* total degree.

*2.3. Realization of multi-resolution methods based on self-loop rescaling*
By the self-loop rescaling strategy, the form of the null model and the weight of it in modularity can be adjusted at one's pleasure. In the following, we derive several multi-resolution methods by the instantiation of the general self-loop rescaling strategy.

(1) Multi-resolution modularity using configuration null model: By setting $k_i^{eff} = k_i$ in Eq.(2), equivalently to provide each node with a self-loop of strength $k_i \cdot (\gamma - 1)$, where $k_i$ is the degree of node $i$ in the network, the multi-resolution modularity can be written as,

$$Q^{CM} = Q\left\{A_{ij} + \gamma k_i \cdot I_{ij} - k_i \cdot I_{ij}\right\}$$
$$= \frac{\gamma - 1}{\gamma} + \frac{1}{\gamma}\sum_s\left(\frac{k_s^{in}}{2M} - \gamma\left(\frac{k_s}{2M}\right)^2\right). \quad (3)$$

As we see, the multi-resolution modularity still preserves the configuration null model (CM) in modularity [8, 43], while adds an important pre-factor of tuning the contribution of the null model. For simplicity, the multi-resolution modularity using the CM null model is denoted by ModCM.

(2) Multi-resolution modularity using the Erdös-Rényi (ER) null model: By setting $k_i^{eff} = \bar{k}$, equivalently to provide each node with a self-loop of strength $(\gamma \cdot \bar{k} - k_i)$, where $\bar{k}$ is the mean degree of nodes in the network, one can obtain the multi-resolution modularity using the Erdös-Rényi (ER) null model,

$$Q^{ER} = Q\{A_{ij} + \gamma \bar{k} \cdot I_{ij} - k_i \cdot I_{ij}\}$$
$$= \frac{\gamma - 1}{\gamma} + \frac{1}{\gamma} \sum_s \left( \frac{k_s^{in}}{2M} - \gamma \left( \frac{\bar{k}_s}{2M} \right)^2 \right). \quad (4)$$

As we see, this method changes the CM null model as the ER null model [8, 43] and adds a pre-factor of tuning the contribution of the null model. For simplicity, the multi-resolution modularity using the ER null model is denoted by ModER.

(3) Multi-resolution modularity using mixing null model: By setting $k_i^{eff} = [(\gamma - 1)\bar{k} + k_i]/\gamma$, one can obtain the multi-resolution modularity,

$$Q^{MIX} = Q\left\{A_{ij} + \gamma \frac{(\gamma - 1)\bar{k} + k_i}{\gamma} \cdot I_{ij} - k_i \cdot I_{ij}\right\}$$
$$= \frac{\gamma - 1}{\gamma} + \frac{1}{\gamma} \sum_s \left( \frac{k_s^{in}}{2M} - \gamma \left( \frac{(k_s + \bar{k}_s(\gamma - 1))/\gamma}{2M} \right)^2 \right). \quad (5)$$

Here the *effective* total degree of group $s$ $k_s^{eff} = (k_s + \bar{k}_s(\gamma - 1))/\gamma$, where $k_s$ is the total degree of group $s$, and $\bar{k}_s = n_s \bar{k}$ ($n_s$ is the number of vertices in group $s$). This method is to simultaneously change the CM null model as a mixing null model (MIX) of CM and ER (the MIX model is to vary with $\gamma$), and add a pre-factor $\gamma$ of tuning the contribution of the null model. For simplicity, the modularity using the MIX null model is denoted by ModMIX. When $\gamma=1$, the MIX null model becomes the CM null model. When $\gamma$ is very large, the MIX null model becomes the ER null model. Thus, the behaviors of ModMIX may be between ModCM and ModER.

**3. Application**

From macro- to micro-scales, i.e., from whole network as one group to the network splitting into a set of separated nodes, the multi-resolution methods can browse community structures at different levels with the increase of their resolution parameters. However, during the intermediate processes of the multi-resolution methods, different methods often give different intermediate structures of networks. Here, we apply the above multi-resolution methods to several synthetic and real networks to investigate the modular structures of the networks (our analysis focuses on un-weighted networks).

*3.1. Synthetic networks*
3.1.1 Fortunato-Barthélemy graph
Firstly, the multi-resolution methods are applied to the graph that consists of two large cliques of 20 nodes and two small cliques of 5 nodes (see figure 1(a)), which was proposed by Fortunato and Barthélemy to demonstrate the resolution limit of the Newman-Girvan modularity [24]. As we know, the optimization of the Newman-Girvan modularity cannot separate the two small cliques, corresponding to the community partition marked by {03} in figure 1(b). The above multi-resolution methods are able to find the expected partition (marked by {04} in figure 1(b)) that consists of four separated cliques by scanning the resolution parameter. It means that the resolution limit of the Newman-Girvan modularity is overcome by the methods in region {04}.

For all the three methods, the plateau region on the left of {03} in figure 1(b) corresponds to the community partition consisting of two groups: the one formed by merging two small cliques into the adjacent large clique, and the other being the remaining clique. On the right of region {04}, for ModCM and ModMIX, we can see a relatively stable plateau region corresponding to a community partition with 42 groups, resulting from the complete breakup of two large cliques and the maintenance of two small cliques. While it is difficult to find this community partition for ModER, as the complete breakup of four cliques seems to appear simultaneously (leading to the network splitting into a set of single-vertex groups). Moreover, all the partitions that are stable at a relatively wide region are of importance. But, the length of the stable plateau regions, such as {03} and {04} in figure 1(b), are different for different methods. And the most stable partitions are also different for different methods. Take the two representative regions {03} and {04} for example, the region {03} seems to be more stable for ModCM in according to the length of the plateau region, while the region {04} seems to be more stable for ModMIX and ModER.

3.1.2 Homogeneous hierarchical graph

Secondly, the multi-resolution methods are applied to the *homogeneous* hierarchical graph with 256 nodes and two predefined hierarchical levels in [45] (see figure 2(a)). The first level consists of 4 groups of 64 nodes and the second level consists of 16 groups of 16 nodes. The number of links of each node with the most internal community is 13, the number of links of each node with the most external community is 4, and the number of links with any other node at random in the network is 1. Figure 2(b) shows that the three multi-resolution methods can easily find the community partitions at the two scales (marked respectively by {04} and {16}) by scanning the resolution parameter. While the optimization of the Newman-Girvan modularity can only identify the community partition at the first level (four groups of 64 nodes), marked by {04} in figure 2(b).

3.1.3 Heterogeneous hierarchical graph

Thirdly, the multi-resolution methods are applied to the *heterogeneous* hierarchical graph with 256 nodes and two predefined internal levels (see figure 1(a) in [46]). The un-weighted edge connection probabilities at each level are the following: Level 2 has $p2 = 0.9$ between nodes in the same group with group sizes from 5 to 22 nodes (has 16 groups), Level 1 has $p1 = 0.3$ between nodes in different groups with merged group sizes from 33 to 76 nodes (has 5 merged groups). (Level 0 corresponds to the whole graph of 256 nodes with $p0 = 0.1$ between nodes in different merged groups.) Similarly, the optimization of Newman-Girvan modularity can only identify the community partition at the first level (5 groups, marked by {05} in figure 3). However, as shown in figure 3, the multi-resolution methods can find the community partitions at the two predefined internal scales (marked respectively by {05} and {16}).

3.1.4 Clique-star graph

Fourthly, we apply the multi-resolution methods to a toy model network which consists of a clique of 10 vertices and four star-like structures of different sizes (see figure 4(a)). As shown in figure 4 (b), the ModER and ModMIX methods can find the expected community partition of the network that consists of four predefined groups (one clique and four stars), but ModCM cannot identify the expected partition in the network. This is because ModCM encounters a kind of limitation in this network—the clique has broken up before small communities (stars) become detectable [27, 28]. According to the results in figure 1(b) and figure 4(b), the tolerance of ModMIX and ModER against the limitation in these networks seems to be stronger than ModCM. Moreover, if we reduce the size of the clique in the toy model, ModCM can escape from the limitation. Therefore, when the community size difference is not very large, all the methods will be able to find the expected partition in the network.

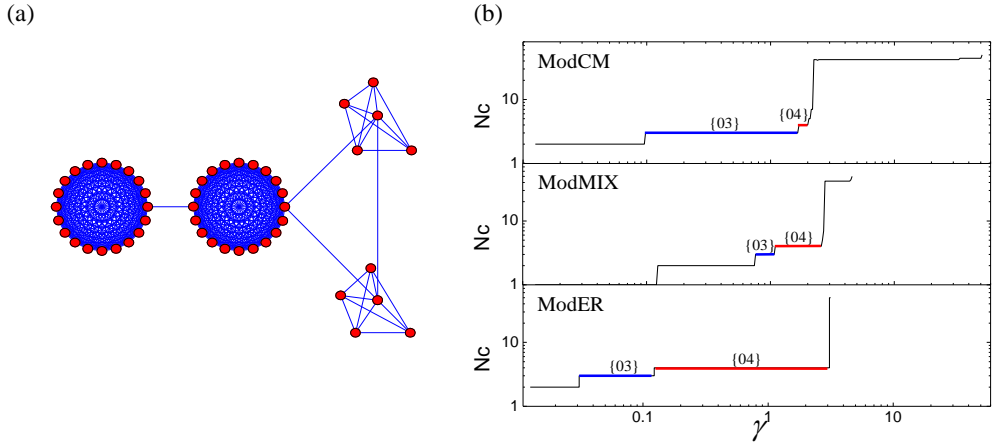

**Figure 1.** (Colour on-line) (a) The graph that consists of two cliques of 20 nodes and two cliques of 5 nodes, which was proposed to demonstrate the resolution limit of modularity in [24]. This limit is overcome at scale {04}(red) providing us with the partition expected. (b) Curves of the $\gamma$-values and the number Nc of groups obtained at the optimal partition for ModCM, ModER, and ModMIX in the graph. The bold solid lines mark the highlighted partitions: {03}(blue) for the partition where two small cliques merge, and {04}(red) for the partition where four predefined cliques are found. To identify the plateaus for the stable partitions, one can also use other statistical measures, such as the normalized mutual information and the variation of information [40]. Other non-highlighted partitions are discussed in the main text.

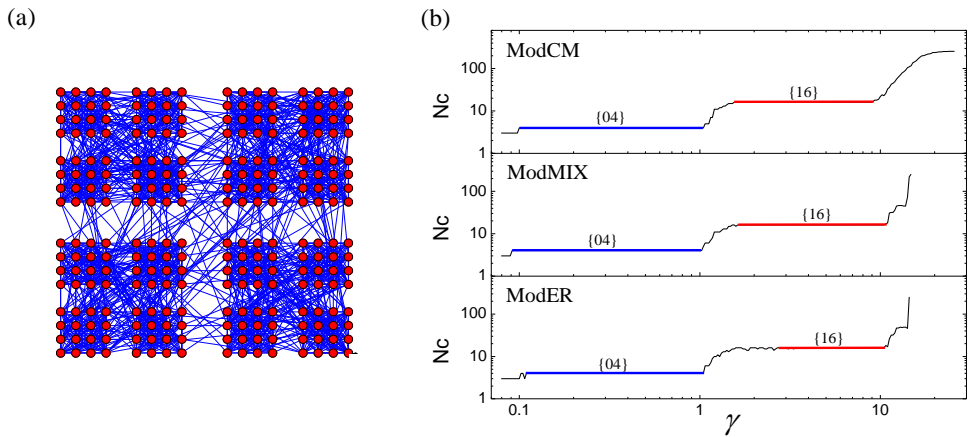

**Figure 2.** (Colour on-line) (a) A homogeneous hierarchical graph with 256 nodes and two predefined hierarchical levels in [45] (see the main text). Both levels are revealed by the methods. (b) Curves of the $\gamma$-values and the number Nc of groups obtained at the optimal partition for ModCM, ModER, and ModMIX in the graph. The bold solid lines mark the two predefined hierarchical levels: {04}(blue) and {16}(red).

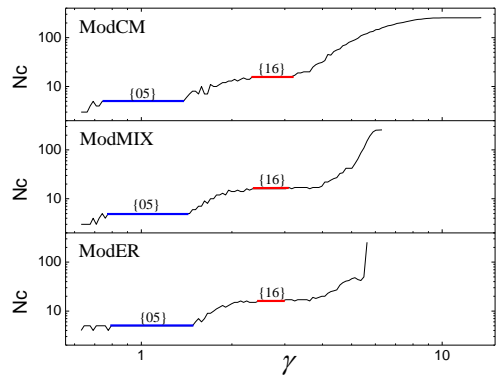

**Figure 3.** (Colour on-line) (a) A heterogeneous hierarchical graph with 256 nodes in [46] (see the main text). The predefined hierarchical levels can be revealed by the methods. (b) Curves of the $\gamma$-values and the number Nc of groups obtained at the optimal partition for ModCM, ModER, and ModMIX in the graph. The bold solid lines mark the two predefined hierarchical levels: {05}(blue) and {16}(red).

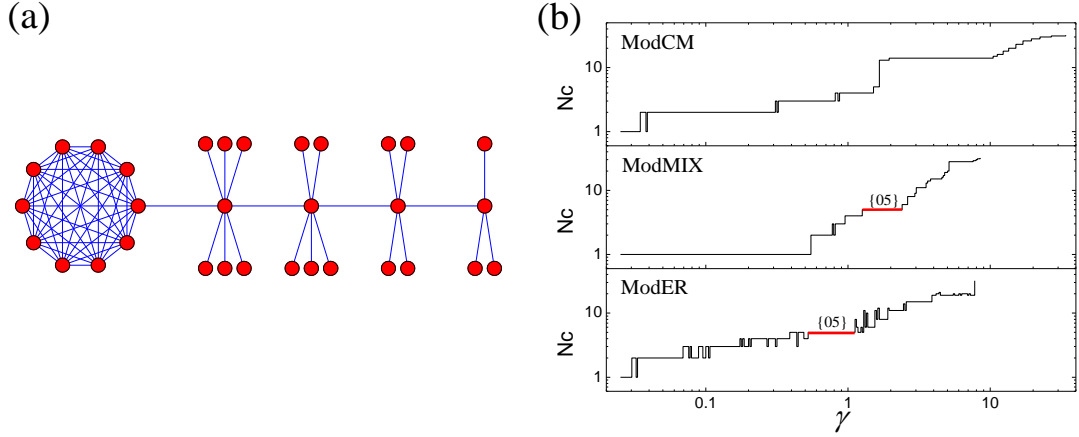

**Figure 4.** (Colour on-line) (a) A toy model network consisting of a clique of 10 vertices and four stars of different sizes. (b) The curve of $\gamma$-value and the number $Nc$ of groups found by ModCM, ModER and ModMIX, in the toy model network with communities of different sizes and densities. The bold solid lines {05}(red) mark the expected partition where four predefined groups (one clique and four stars) are found. ModCM cannot identify the expected partition in the network.

*3.2. Real-world networks*

Here, we apply the multi-resolution methods to two real-world networks: the karate club network of Zachary [47] and the dolphin social network of Lusseau et al [48, 49], which are two well-known networks for testing community detection algorithms. The two networks both were split into perfectly identifiable parts during study of researchers (see figure 5(a) and figure 6(a)). An ideal community detection method should be able to identify these modular structures without a priori knowledge of these parts.

By the optimization of the Newman-Girvan modularity, one often finds the community partition with four groups for the karate network and the community partition with five groups for the dolphin network, while not the real splits of the networks. Figures 5(b) and 6(b) show that the multi-resolution methods can find the community partitions representing the real splits of the two networks without misclassification of any node, corresponding to the stable plateau regions denoted by {02}.

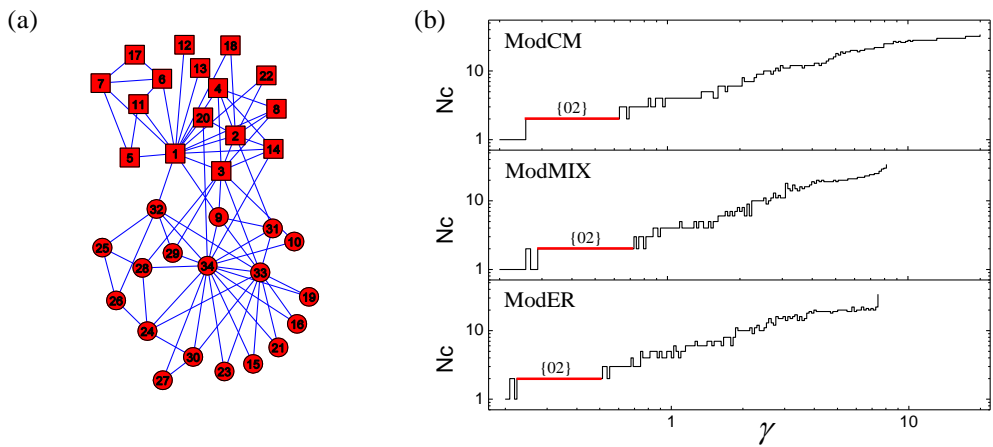

**Figure 5.** (Colour on-line) (a) The network of friendship relations between 34 members in the karate club study of Zachary at a US university in 1970. The network split into two groups during Zachary's study, due to the disagreement between the administrator and the instructor of the club (see squares and circles). (b) Curves of the $\gamma$-values and the number Nc of groups obtained at the optimal partition by different methods in the network. The (red) bold solid lines {02} mark the community partition corresponding to the real split of the network by the methods.

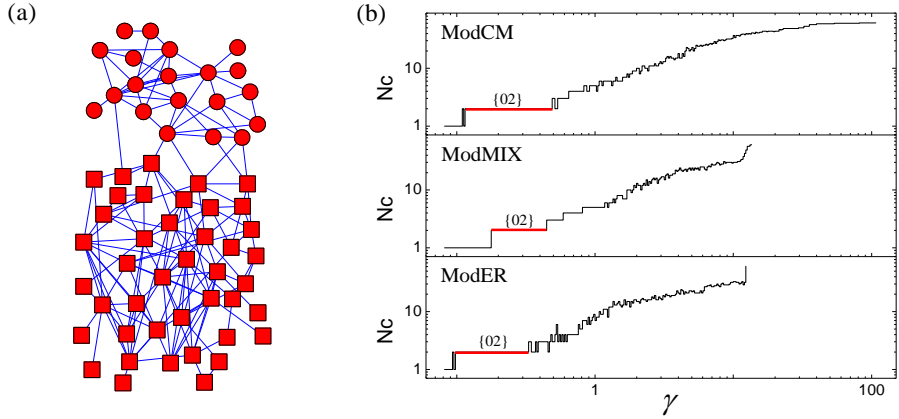

**Figure 6.** (Colour on-line) (a) The dolphin social network of Lusseau et al [48, 49], constructed from observations of a community of 62 bottle-nose dolphins over a period of 7 years from 1994 to 2001. During the study of the observations, the dolphin network was divided into two identifiable parts (see squares and circles), due to the temporary disappearance of a dolphin (SN100). (b) Curves of the $\gamma$-values and the number Nc of groups obtained by different methods in the network. The (red) bold solid lines {02} mark the community partition corresponding to the real split of the network by the methods.

## 4. Discussion

By generalizing the self-loop rescaling strategy, we have proposed a more general and uniform ansatz for the design of multi-resolution methods based on self-loop rescaling, by which one can construct many different (multi-resolution) modularity methods with any expected null models. We have constructed and validated three multi-resolution modularity methods of this type, though, in principle, more complicated multi-resolution modularity can be obtained by the self-loop rescaling strategy. The AFG method is only a special case of this uniform ansatz [25]. Differently from the AFG method, ModMIX is to provide each node with a self-loop that is dependent of the mean degree of network under study; while the AFG method is to assign to each vertex a self-loop directly by a tunable parameter (which is independent of the structures of networks).

We find that some important quality functions for community detection can be unified in the framework of the self-loop rescaling strategy [8, 25, 27, 28, 41, 46]. That is, the equivalent target functions of these different quality functions can be derived uniformly by the self-loop rescaling strategy from the *Newman-Girvan modularity* function. It is worth noting that the Potts model has been obtained extensive application and discussions in community detection [8, 40, 41, 46, 50]. Each of Hamiltonian based on Potts model can find its counterpart of modularity (corresponding to the negative of the corresponding modularity). The maximum of modularity is reached when Hamiltonian is minimal, and maximizing the modularity is hence equivalent to finding the spin configuration that minimizes the Hamiltonian. We find that, for example, the equivalent quality functions of the RN method [46] and the TD method [41] can be given by ModER respectively with $\gamma_{ER} = \gamma_{RN}/(p + p\gamma_{RN})$ and $\gamma_{ER} = \gamma_{TD}/p$, where $p = \bar{k}^2/2M$. Thus the application of the self-loop rescaling strategy may be extended to these studies of the Hamiltonians based on Potts model, helping the understanding of the related methods. Moreover, it is shown that the self-loop rescaling, when properly defined, does not affect the spectral properties of the networks[51].

Because it is only indirectly that the self-loop rescaling strategy modifies the *Newman-Girvan* modularity, all the resulting multi-resolution modularity by the self-loop rescaling strategy can be optimized directly by the existing algorithms for optimizing *Newman-Girvan modularity* with minimum code development, though some quality functions, such as the Hamiltonian functions in the RN method and the TD method, may need some skillful parameter transformations. Moreover, in literatures, many effective algorithms were proposed for the optimization of the *Newman-Girvan* modularity. By using the general self-loop rescaling strategy, these optimization algorithms (including the codes of the algorithms) can be re-used in the optimization of various quality functions for community detection as well as in the studies of the multi-scale structures of networks. In brief, by the general self-loop rescaling strategy, various quality functions for community detection have more choices for the optimization algorithms, but also the application of many existing modularity optimization algorithms can be extended in community detection.

Moreover, some methods or measures in literatures were proposed for identifying the most relevant resolutions or the best partitions for the networks. For example, Ronhovde et al used the information comparisons among independent solutions to quantitatively measure the best resolutions [40]. It is a topic of interest to identify the most relevant resolutions, while it is out of the scope of the paper. As in reference [25], we determine the relevant community partitions in terms of the stability of community partitions against the change of resolution parameters, i.e. the length of the stable plateaus.

## 5. Conclusion

In recent years, many multi-resolution methods have been proposed to discover community structures at different scales, based on various approaches. In this paper, we presented a type of multi-resolution methods by using the generalized self-loop rescaling strategy. These multi-resolution methods derived by using the rescaling strategy, as typical examples of the self-loop rescaling strategy, were applied to several synthetic and real-world networks. The results show that these methods can find the pre-defined substructures in the synthetic networks and exact splits of the real-world networks reported in the literatures.

The self-loop rescaling strategy provides one uniform ansatz for the design of these multi-resolution methods. Some existing quality functions can find the counterparts of modularity based on the self-loop rescaling strategy by suitable parameter transformations. The self-loop rescaling strategy can indirectly modify the definition of modularity function by providing each vertex with a suitable self-loop, and thus, the resulting multi-resolution modularity can be optimized directly by the existing modularity optimization algorithms with minimum code development. This can enrich the choices of the optimization algorithms for optimizing these quality functions, but also extends the application of the existing modularity optimization algorithms in community detection. We hope that the study in the paper can help further understand the multi-resolution methods in community detection and provide useful insight into designing new community detection methods.

**Acknowledgement**

We sincerely appreciate the careful and professional review of the referees as well as the valuable comments for the improvement of the paper. This work has been supported by the Scientific Research Fund of Education Department of Hunan Province (Grant Nos. 14C0126, 11B128, 14C0127, 14C0112, 12C0505 and 14B024) and Hunan Provincial Natural Science Foundation of China (Grant No. 13JJ4045), and partly by the Doctor Startup Project of Xiangtan University (Grant No. 10QDZ20).